\documentclass[twocolumn]{jpsj2} %% two-column layout
\title{Krylov Subspace Method for Molecular Dynamics Simulation \\
based on Large-Scale Electronic Structure Theory
}

\author{Ryu \textsc{Takayama}$^{1,2}$\thanks{E-mail address: takayama@coral.t.u-tokyo.ac.jp}, 
        Takeo \textsc{Hoshi}$^{2}$ and Takeo \textsc{Fujiwara}$^{2}$ }

\inst{$^{1}$ Research and Development for Applying Advanced Computational Science and Technology, \\
Japan Science and Technology Agency, 4-1-8 Honcho, Kawaguchi-shi, Saitama 332-0012, Japan\\
$^{2}$ Department of Applied Physics, University of Tokyo, 7-3-1 Hongo, 
Bunkyo-ku, Tokyo 113-8656, Japan}

\abst{
For large scale electronic structure calculation, 
the Krylov subspace method is introduced to calculate the one-body density matrix 
instead of the eigenstates of given Hamiltonian. 
This method provides an efficient way to extract the essential character 
of the Hamiltonian within a limited number of basis set. 
Its validation is confirmed by the convergence property of the density matrix 
within the subspace. 
The following quantities are calculated; 
energy, force, density of states, and energy spectrum. 
Molecular dynamics simulation of Si(001) surface reconstruction is examined as an example, 
and the results reproduce the mechanism of asymmetric surface dimer. 
}

\kword{Krylov subspace, large scale electronic structure calculations, 
density matrix, 
tight-binding molecular dynamics, 
surface reconstruction, parallel computation, 
hybrid scheme within quantum mechanics
}

\begin{document}
\maketitle

%%%%%%%%%%%%%%%%%%%%%%%%%%%%%%%%%%%%%%%%%%%%%%%%%%%%%%%%%%%
\section{Introduction}
\label{sec:intro}
The study of nanoscale systems requires 
a large-scale atomistic simulations with quantum mechanical freedoms of electrons.
The practical requirement to carry out the simulations is 
how to extract desired quantities 
from a given large Hamiltonian matrix, not only accurately 
but also efficiently. 
Simulation methods in large scale systems 
have been studied already in the last decade.~\cite{Ordejon98a,Galli.00,Wu02}
In order to execute molecular dynamics simulation, 
one needs information about 
the total energy and forces on an individual atom, 
and these physical quantities should be obtained by means of either 
eigen states $\mid\phi_{\alpha}\rangle$ or the 
one-body density matrix $\rho$  of the system;
\begin{eqnarray}
\rho =
\sum_{\alpha}^{} \mid\phi_{\alpha} \rangle \langle \phi_{\alpha} \mid
 f \Bigl(\frac{\varepsilon_{\alpha}-\mu}{k_{\rm B}T}\Bigr) .
\end{eqnarray}
Here $f \Bigl(\frac{\varepsilon_{\alpha}-\mu}{k_{\rm B}T}\Bigr) $ 
is the Fermi-Dirac distribution function as a function of the eigen energy $\varepsilon_{\alpha}$ of 
the eigen states $\mid\phi_{\alpha}\rangle$ and the chemical potential $\mu$ of the system.

The molecular dynamics calculation in large-scale systems 
can be  done on the basis of  transferable short-range 
tight-binding Hamiltonians $H$, 
where we  calculate the physical property $\langle X \rangle $ as  
\begin{equation}
\langle X \rangle ={\rm Tr}[\rho X] = \sum_{ij} \rho_{ij} X_{ji} . 
\label{eq:physica-quantity}
\end{equation}
Here $i$ and $j$ are suffices of atomic site and orbital. 
The energy and forces acting on an atom are contributed 
only by elements that
have  non-zero values of the Hamiltonian matrix.         
In other words, even though the density matrix is of long range,  
only the short range behavior of the density matrix is essential
\cite{Kohn96}.
Therefore, the essential methodology for large scale calculations is 
how to obtain the short range part of  
the density matrix ${ \rho}$ without 
calculating eigen states of the original Hamiltonian. 
The essential point here is the fact that we adopt the short-range tight-binding 
Hamiltonian and this makes  computation local. 
We will only comment here that the short-range tight-binding Hamiltonian 
can be always constructed both in insulators and metals 
from the first principle theory.~\cite{OKAndersen.84,OKAndersen.00}

We have developed a set of methods, without calculating eigenstates, 
of large-scale atomistic simulations, which are  based 
on generalized Wannier state and hybrid scheme 
within fully quantum mechanical description of electron 
systems \cite{hoshi-fujiwara.00,hoshi-fujiwara.01,hoshi-fujiwara.03,hoshi.03,geshi-etal.03}. 
These methods are rigorously a linear scale simulation in atom number, 
and were tested  upto 10$^6$ atoms by using a standard workstation. 
The generalized Wannier states 
are defined formally as unitary transformation 
of the occupied eigen states, 
though eigen states are not actually obtained. 
This method is practical and efficient in covalent bonded materials, 
where  the localized Wannier states  reproduce well 
the electronic structure energy and
the density matrix, at least  its short-range behavior. 
We observed that 
the bond forming and breaking processes are well described 
in the localized Wannier states as changes  
between a bonding and non-bonding 
orbital~\cite{hoshi-fujiwara.03,hoshi.03}. 
In metallic systems, however, situations are quite different 
and other practical methods should be developed.

The aim of the present work is 
to establish an novel extension  of methodology  
%with reliable processes and also being 
practical in  metals. 
We will develop a novel method based 
on the Krylov subspace (KS) method 
to achieve computational efficiency.  
In \S \ref{sec:theory} we review the KS method 
and the density matrix is represented in the KS. 
An example will be presented based on our numerical results. 
These include a discussion of locality of off-diagonal elements of 
the density matrix. 
In \S \ref{sec:results}, as an example of the molecular dynamics 
simulation, the reconstruction of Si (001) surface 
will be discussed.  
We will also show how the energy spectrum can be obtained in our 
developed method. 
In \S \ref{sec:conclusion} we summarize the work presented in this paper.

%%%%%%%%%%%%%%%%%%%%%%%%%%%%%%%%%%%%%%%%%%%%%%%%%
\section{Density Matrix Calculation based on Krylov Subspace Method}
\label{sec:theory}
In this section, we will show theoretical background of the 
KS (Krylov subspace) method 
to extract density matrix for molecular dynamics simulation. 
Short review of the KS method is followed by analysis of 
its arithmetic structure 
including convergence property which justifies the present method.

%%%%%%%%%%%%%%%%
\subsection{Krylov subspace method~\cite{Golub-VanLoan.89,Vorst.03}}
The KS (Krylov subspace) method gives the mathematical foundation of 
many numerical iterative algorithms such as the conjugate gradient method. 
%This method is based on an iterative method and 
%to replace a given system by some nearby system that can be more easily solved. This method 
This method 
provides an efficient way to extract the essential character 
of the original Hamiltonian within a limited number of basis set. 
%
%During the iterative procedure the evaluation of the method focuses 
%on how quickly the iterate converges, 
%Since this method is based on the iterative method 
%and the convergence depends on the sparsity of the original Hamiltonian and 
%the choice of the starting vector. 
%
Starting from a certain vector $\mid i \rangle$, 
a subspace of the original Hilbert space is generated by a set of vectors; 
\begin{equation}
\label{eq:krylov0}
\mid i \rangle, \ \  H\mid i \rangle, \ \  H^2\mid i \rangle, \ \  \cdots, \  H^{\nu_{\rm K}-1}\mid i \rangle .
\end{equation}
The subspace spanned by the basis vectors $\{H^n\mid i\rangle \}$  
in eq.~(\ref{eq:krylov0}) 
is generally called the  Krylov subspace (KS)  
in the mathematical textbooks. 
The dimension of the KS is denoted as $\nu_{\rm K}$. 
We will denote the orthonormalized basis vectors in the KS as 
\begin{equation}
\label{eq:krylov1}
\mid K_{1}^{(i)} \rangle (\equiv \mid i \rangle), \ \  \mid K_{2}^{(i)} \rangle, \ \  \mid K_{3}^{(i)} \rangle, 
\ \  \cdots, \  \mid K_{\nu_{\rm K}}^{(i)} \rangle .
\end{equation}
Since the matrix $H$ is Hermitian, the Gram-Schmidt orthonormalization procedure gives one possible 
(but not necessary) choice of the basis set that satisfies the  
three-term recurrence relation called the Lanczos process; 
\begin{equation} 
b_n\mid K_{n+1}^{(i)}\rangle = (H- a_n) \mid K_n^{(i)}\rangle - b_{n-1}^{\ast}\mid K_{n-1}^{(i)}\rangle ,                    
\label{recur}
\end{equation} 
with $b_{-1}\equiv 0$. 
%This process is called the Lanczos process. 
Hereafter we restrict ourselves to real symmetric Hamiltonian matrix, $H$.

%We will see that the procedure generating the KS can be terminated 
%in a finite number of steps $\nu_{\rm K}$ as a reliable approximation. 
%
%As an important character of the KS method, 
%information about $H$'s external eigenvalues 
%tends to emerge long before the procedure is completed. 
%So the procedure can be terminated in a finite number of steps. 
%

%due to the arithmetic structure of the KS method
%With an appropriate choice of the starting vector the convergence of the 
%iterative process improves and then reduces $\nu_{\rm K}$. 
%
From the practical point of view of calculations, 
the procedure of matrix-vector multiplication, $H\mid K_n^{(i)}\rangle$, consumes the CPU time mostly, 
%the procedure to create $H^{n}\mid i\rangle$  consumes the CPU time mostly, 
then the number of bases in the KS  ($\nu_{\rm K}$) should be 
chosen to be much smaller 
than that of the original Hamiltonian matrix. 
This drastic reduction of the matrix size or the dimension of the KS 
is a great advantage for a practical large-scale calculations. 
The dimension of the KS  $\nu_{\rm K}$ should be  chosen, 
for example, as $\nu_{\rm K}=30$. 
We then denote the reduced Hamiltonian as $H^{K(i)}$ for the KS 
$\{ \mid K_n^{(i)} \rangle \}$.

%\begin{eqnarray}
%G_{ij} &=& \langle i \mid G \mid j \rangle = \langle i \mid u G^{K(i)} u^t \mid j \rangle 
%\\
%%
%&=& \sum_{mn} \langle i \mid K_{im} G_{mn}^{K(i)} (\ ^{t}u)_{nj} \mid j \rangle 
%\\
%%
%&=& \sum_{n} G_{1n}^{K(i)} (\ ^{t}u)_{nj} 
%\label{eq:G-recursSLr}
%%
%\end{eqnarray}

%%%%%%%%%%%
\subsection{Density matrix calculation in the Krylov subspace}
In order to extract desired density matrix, we diagonalize the reduced Hamiltonian matrix $H^{K(i)}$. 
Once one obtains the eigenvalue $\varepsilon^{(i)}_{\alpha}$ and eigenvector $\mid w^{(i)}_{\alpha} \rangle$ as
\begin{eqnarray}
H^{K(i)} \mid w^{(i)}_{\alpha} \rangle = \varepsilon^{(i)}_{\alpha} \mid w^{(i)}_{\alpha} \rangle ,
\label{eq:eigeneq}
\end{eqnarray}
the eigen vector  can be expanded in terms of  the basis  $\mid K_{n}^{(i)} \rangle$; 
\begin{eqnarray}
\mid w^{(i)}_{\alpha} \rangle & = & \sum_{n=1}^{\nu_{\rm K}} C_{\alpha n}^{\ast} \mid K_{n}^{(i)}\rangle .
\label{eq:decomposition}
\end{eqnarray}
We introduce the density matrix operator  within the KS:  
\begin{eqnarray}
\hat{\rho}_{}^{K(i)}
&\equiv&
\sum_{\alpha}^{\nu_{\rm K}} \mid w^{(i)}_{\alpha} \rangle \langle w^{(i)}_{\alpha} \mid 
f\left(\frac{\varepsilon^{(i)}_{\alpha} - \mu}{k_{\rm B}T}\right).
\label{dm-KS}
\end{eqnarray}
%

%This will be exact, $\rho^{K(i)} = \rho$, when $\nu_{\rm K}$ is equal to the 
%matrix size. 
%Therefore, this method is an iterative method that gives a sequence of 
%$\rho^{K(i)}$ as the function of $\nu_{\rm K}$. 
%The evaluation of a iterative method invariably focuses on how 
%quickly the iterate converges. 
%We will see the dimension of the KS  $\nu_{\rm K}$ can be  reduced, 
%for example, as $\nu_{\rm K}=30$. 

%The only one approximation in 
The essence of 
the present method is the replacement of the density matrix 
$\langle i \mid \hat{\rho} \mid j \rangle $ by that of the KS 
$ \langle i \mid \hat{\rho}^{K(i)} \mid j \rangle $;
\begin{equation}
\langle i \mid \hat{\rho} \mid j \rangle \Rightarrow  \langle i \mid \hat{\rho}^{K(i)} \mid j \rangle. \  \ 
%{\rm  or} \ \ 
% \hat{\rho}  \Rightarrow   \hat{\rho}^{K(i)}  \ .
\label{appr}
\end{equation}
Once this procedure is allowed, it is a great advantage from the 
view point of practical calculations.

Let us introduce the  projection operator; 
\begin{eqnarray}
\hat{P}^{K(i)}
&\equiv&
\sum_{\alpha}^{\nu_{\rm K}} \mid w^{(i)}_{\alpha} \rangle \langle w^{(i)}_{\alpha} \mid 
=\sum_{n}^{\nu_{\rm K}} \mid K_{n}^{(i)} \rangle \langle K_{n}^{(i)} \mid .
\end{eqnarray}
The crucial point is for the calculation of 
$\langle i \mid \hat{\rho} \mid j \rangle$ 
that, though the state $\mid i \rangle$ is an element of the KS  
($P^{K(i)}\mid i \rangle =\mid i \rangle$), 
the state $\mid j \rangle$ may be not an element completely included within 
the KS  ($P^{K(i)}\mid j \rangle \ne \mid j \rangle$ nor $0$).
Even so, the density matrix of the KS 
$ \langle i \mid \hat{\rho}^{K(i)} \mid j \rangle $  holds the following 
relation;
\begin{eqnarray}
\langle i \mid \hat{\rho}^{K(i)} \mid j \rangle 
%&=& \langle i \mid \hat{P}^{K(i)} \hat{\rho}_{}^{K(i)} \mid j \rangle
%\label{eq:project1}
%\\
%&=& \langle i \mid \hat{\rho}_{}^{K(i)} \hat{P}^{K(i)} \mid j \rangle
%\label{eq:project2}
%\\
%&=& 
= \sum_{n}^{\nu_{\rm K}} \langle i \mid \hat{\rho}_{}^{K(i)} \mid K_{n}^{(i)} \rangle
\langle K_{n}^{(i)} \mid j \rangle. 
\label{eq:project3}
\end{eqnarray}
%
%In eq.(\ref{eq:project1}), we use the relation  
%
%\begin{eqnarray}
%\hat{P}^{K(i)} \mid i \rangle &=& \mid i \rangle,  \\
%\hat{P}^{K(i)} \mid j \rangle & \neq & \mid j \rangle. \hspace{1.5cm} (i \neq j)
%\end{eqnarray}
%
To show eq.~(\ref{eq:project3})
we use a  relation 
\begin{equation}
\hat{\rho}_{}^{K(i)}=\hat{\rho}_{}^{K(i)} \hat{P}^{K(i)}. 
\end{equation}
%
%\begin{equation}
%[\hat{\rho}_{}^{K(i)}, \hat{P}^{K(i)} ] = 0. 
%\end{equation}
%

%Since these quantities are composed of the eigen states of 
%the Hamiltonian, $\mid w^{(i)}_{\alpha} \rangle$. 
%
The replacement eq.~(\ref{appr}) is rigorous 
when $\nu_{\rm K}$ is equal to the dimension of the original Hilbert space. 
When  $\nu_{\rm K}$ is much smaller, 
this replacement  (\ref{appr}) can be justified 
only if  the convergence of the summation 
in eq.~(\ref{eq:project3}) is 
fast enough and the contribution from large $n$ is negligible. 
We will check the $n$ dependence of 
both $\langle i \mid \hat{\rho}_{}^{K(i)} \mid K_{n}^{(i)} \rangle$ and $\langle K_{n}^{(i)} \mid j \rangle$ 
in Subsection \ref{subsec:Convergence}.

Considering the spin degeneracy, the relation of electron number $N_{\rm elec}$ 
and the chemical potential 
$\mu$ can be given as 
\begin{eqnarray}
\frac{N_{\rm elec}}{2} &=& \sum_i \langle i \mid \hat{\rho}^{K(i)} \mid i \rangle 
\nonumber \\
&=&  \sum_{i \alpha} \mid \langle i \mid w^{(i)}_\alpha \rangle \mid^2
f\left(\frac{\varepsilon^{(i)}_{\alpha} - \mu}{k_{\rm B}T}\right),
\label{chemPot}
\end{eqnarray}
which is used to determine the chemical potential $\mu$ 
in the system.

For short summary of this subsection, 
we note that the 
%approximation 
essential 
procedure is only the part 
reducing the dimension of the original Hamiltonian matrix $H$ 
to that of the reduced matrix $H^{K(i)}$. 
Once we obtain $\{ \mid w^{(i)}_{\alpha} \rangle \}$ 
the cost of calculating 
$\{ \langle i \mid \rho^{K(i)} \mid j \rangle \}$ for 
necessary enough number of neighboring sites and orbitals $j$ of 
a fixed $i$ is of the order of  one, 
independent of the system size or the total number of atoms.
And, furthermore, the calculation of them is perfectly parallelizable 
with respect to  sites and orbitals $i$. 
%

%---------------------------------------
\begin{figure}[htbp]
%\begin{center}
\resizebox{0.48\textwidth}{!}{
%  \includegraphics{./DECAY01.eps}
% SLD24a512/figures/DECAY01.ps
  \includegraphics{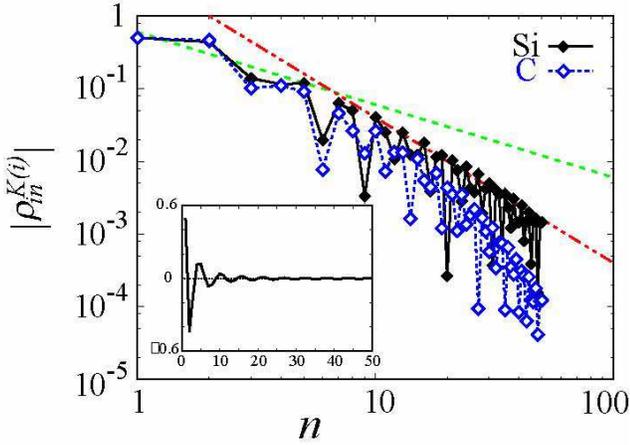}
% SGI-JAERI/SLD24a4096larger/testOpenMP2/figures/DECAY06.ps
}
%\end{center}
\caption{ 
Decay properties of the off-diagonal density matrix, 
$\rho_{in}^{K(i)}$, of 262,144 atom for Si (solid square with line) 
and that for C (open square with line) 
as a function of $n$ in the summation of eq.~(\ref{eq:project3}) 
representing number of hoppings from center atom. 
%and that for one dimensional chain model with constant 
%$a$ and $b$ (open circle). 
The dashed line (two dot dashed line) is a guide 
to the eye representing $1/n~ (1/n^2)$ behavior. 
Inset shows $\rho_{in}^{K(i)}$ for Si in linear scale. 
}
\label{fig:decay}
\end{figure}
%---------------------------------------

%%%%%%%%%%%%%%%
\subsection{Convergence properties of the density matrix}
\label{subsec:Convergence}

In order to demonstrate the validity of the replacement eq.~(\ref{appr}), 
we check the convergence of eq.~(\ref{eq:project3}). 
The convergence varies according to the locality of the original Hamiltonian 
as well as the choice of starting basis orbitals. 
To demonstrate the property, we choose a crystal of diamond structure with 262,144 atoms 
using a transferable tight-binding Hamiltonian~\cite{kwon-etal.94}. 
We also choose the four sp$^3$ orbitals per atom 
as starting basis orbitals $\{ | i \rangle \}$.  

Figure~\ref{fig:decay} shows decay property of 
$\rho_{in}^{K(i)} (\equiv \langle i \mid \hat{\rho}_{}^{K(i)} \mid K_{n}^{(i)} \rangle)$ 
as a function of the $n$ in the summation of eq.~(\ref{eq:project3}). 
As shown in the inset, $\rho_{in}^{K(i)}$ decays oscillatery. 
We plot its absolute value to see the decay behavior. 
In case of silicon, we can read that $\rho_{in}^{K(i)}$ decays as fast as $1/n^2$. 
On the other hand the value of $\langle K_{n}^{(i)} \mid j \rangle$ 
also decays as a function of $n$ 
due to the fact that
the state $\mid K_n^{(i)}\rangle$ extends over sites reached 
with $n$-steps from the starting state $\mid K_1^{(i)}\rangle (\equiv \mid i \rangle)$ 
to cover another localized basis $| j \rangle$ on one site.
Decay property of $\langle K_{n}^{(i)} \mid j \rangle$ 
depends on $j$ but its maximum value in the present system decays as $1/n$ (not shown in the figure). 
Therefore, the products in eq.~(\ref{eq:project3}) decays as $1/n^3$. 
We examined several cases in different system size (512, 4096, 32768, and 262144 atoms), 
and found that the decay property is almost independent of the system size. 
In case of carbon\cite{NOTE-TB}, on the other hand, 
the decay rate of $\rho_{in}^{K(i)}$ is even more faster, 
which can be understood from the locality of 
the Wannier state. \cite{NOTE-locality}
%formed by sp$^3$ orbitals is 
%more localized.~\cite{hoshi-fujiwara.00,hoshi-fujiwara.01,NOTE-locality}  
%We note here that the locality in the present context is of hopping range, $n$. 
%This means the locality in normalized length by lattice constant. 

Since the choice of the starting basis is arbitrary, 
we can choose the four atomic orbitals at each atom site, 
(s, p$_x$, p$_y$, p$_z$) , as starting basis orbitals . 
Here, however, 
we choose the starting bases
$| i \rangle \equiv \mid K_1^{(i)}\rangle$
as the four sp$^3$ orbitals,
because the cohesive mechanism is clarified
with such hybridized bases.
Due to the crystalline symmetry of diamond structure, 
the four sp$^3$ bases are equivalent and only one example 
is enough for the explanation of the cohesive mechanism. 
The dominant interaction in the Hamiltonian is 
the hopping along the sp$^3$ bond. 
If we ignore other hoppings in the Hamiltonian,
%and choose a certain sp$^3$ orbital as a starting basis, 
the Krylov subspace with $\nu_k=2$ gives 
the sp$^3$ bonding and anti-bonding orbitals as 
\begin{eqnarray}
\frac{\mid K_1^{(i)} \rangle  \pm \mid K_2^{(i)}\rangle}{\sqrt{2}},
\end{eqnarray}   
which forms a desirable basis set in the present case.
%This explains why the small number bases in the Krylov subspace 
%gives the correct description of the cohesive mechanism.

%We note here that 
%%%% choice of orbitals %%%
%the choice of the starting bases as sp$^3$ orbitals is crucial 
%to make the dimension $\nu_{\rm K}$ of the KS smaller and the convergence faster. 
%For diamond structure solids, 
%the dominant interaction in the Hamiltonian is 
%the hopping along the sp$^3$ bond. 
%If we ignore other hoppings in the Hamiltonian and 
%choose a certain sp$^3$ orbital as a starting basis, 
%the Krylov subspace with $\nu_k=2$ gives 
%the sp$^3$ bonding and anti-bonding orbitals as 
%\begin{eqnarray}
%\frac{\mid K_1^{(i)} \rangle  \pm \mid K_2^{(i)}\rangle}{\sqrt{2}},
%\end{eqnarray}   
%which forms a desirable basis set in the present case.
%Since the choice of the starting basis is arbitrary, 
%we can choose the four atomic orbitals, 
%(s, p$_x$, p$_y$, p$_z$), as starting basis orbitals. 
%In such a case, however, we need sufficiently large number of steps of 
%operating $H$ in eq.~(\ref{eq:krylov0}) 
%to produce sp$^3$ orbitals on each atomic site. 
%Therefore, the dimension of the KS could be much larger in order to obtain 
%the similar results of the proceeding results. 
%
%Since the four sp$^3$ orbitals per atom 
%are equivalent to the atomic orbitals
%(s, p$_x$, p$_y$, p$_z$), 
%the above choice of the starting bases 
%dose not cause any error.  
%In general, we can choose 
%starting bases for each atom, independently,
%as a technique for faster computational performance.

%%%%%%%%%%%%%%%%%%%%%%%%%%%%%%%%%%

We would consider 
an example of possible slowest convergence of eq.~(\ref{eq:project3}) 
where we can define the Fermi wave vector $k_{\rm F}$; 
Since the three-term recurrence  relation 
in eq.~(\ref{recur}) 
suggests a mapping of the original system  
to a one-dimensional chain model, 
it is instructive to compare with simple consideration of one dimensional 
system with constant energy $a$ and hopping $b$ in eq.~(\ref{recur}). 
This case corresponds to the one-dimensional free space,
in the continuum limit,
and the density matrix is given by analytic form 
\begin{equation}
\rho(x,x') \equiv 
\int_{-k_{\rm F}}^{k_{\rm F}} {\rm e}^{{\rm i}k(x-x')} {\rm d}k = \frac{\sin k_{\rm F} (x-x')}{x-x'}.
\label{eq:analytic-rho}
\end{equation}
This can be understood as $1/n$ behaviour of $\rho_{in}$ with oscillation. 
Even in this case $\langle K_n^{(i)}\mid j\rangle$ 
decays as $1/n$ and the products in eq.~(\ref{eq:project3}) decays as $1/n^2$. 
Though the analysis for other realistic systems like simple cubic lattice 
will be shown elsewhere, 
we should mention that there are several practical examples where  
$\rho_{in}^{K(i)}$ decays as $1/n$.

Further, from the view point of practical calculations, 
the decay rate can be controlled by the temperature factor $k_{\rm B} T$; 
The higher the temperature, the faster the decay \cite{note-Gibbs.Osci}.  
These facts validate the convergence of the summation in eq.~(\ref{eq:project3}) 
and justifies the 
replacement 
%present approximation 
eq.~(\ref{appr}).
%In any case,
%the products in eq.~(\ref{eq:project3}) decay faster than $1/n^2$.  

%%%%%%%%%%%%%%%%
%---------------------------------------
\begin{figure}[htbp]
%\begin{center}
\resizebox{0.48\textwidth}{!}{
  \includegraphics{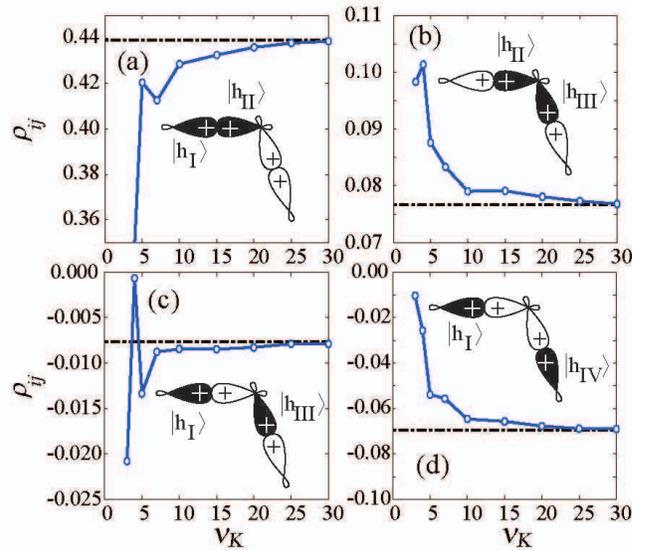}
% BL21/periodic512/figures/RECdep11p.ps
}
%\end{center}
\caption{ 
Reduced matrix size dependence of the off-diagonal elements of the density matrix, $\rho_{ij}$, 
of 512 Si atom (open circle with line). 
The dot dashed line represents the results of exact diagonalization of the original Hamiltonian. 
Inset shows schematic pictures of the sp$^3$ hybrid orbitals 
$\mid {\rm h_{I}} \rangle, \mid {\rm h_{II}} \rangle, \mid {\rm h_{III}} \rangle$, and $\mid {\rm h_{IV}} \rangle$. 
Solid orbitals in each figure represent the combination to contribute the density matrix. 
}
\label{fig:off-density-matrix}
\end{figure}
%---------------------------------------

\subsection{Convergence properties of off-diagonal elements of the density matrix and the total energy}
\label{sec:rhoij-dep}

As an example of the present method, 
we show the calculated density matrix and compare with that of diagonalization 
of the original Hamiltonian. 
We pick up two nearest neighbor bond sites along a linear path 
with four sp$^3$ hybrid orbitals 
$\mid {\rm h_{I}} \rangle$, 
$\mid {\rm h_{II}} \rangle$, 
$\mid {\rm h_{III}} \rangle$, 
$\mid {\rm h_{IV}} \rangle$, 
where two orbitals 
$\{ \mid {\rm h_{I}} \rangle$ and $\mid {\rm h_{II}} \rangle \} $ and 
$\{ \mid {\rm h_{III}} \rangle$ and $\mid {\rm h_{IV}} \rangle \}$ are 
on the same bond sites 
and $\{ \mid {\rm h_{II}} \rangle$ and $\mid {\rm h_{III}} \rangle \} $ are 
on the same atom. 
See inset of Fig.~\ref{fig:off-density-matrix} for the configuration and 
the phases of respective hybrid orbitals. 
The  exact values of these matrix elements are calculated 
by the exact diagonalization of the original Hamiltonian as follows; 
$ \langle {\rm h_{I} }\mid \rho \mid {\rm h_{II} } \rangle = 0.439$, 
$ \langle {\rm h_{II}}\mid \rho \mid {\rm h_{III}} \rangle = 0.078$, 
$ \langle {\rm h_{I} }\mid \rho \mid {\rm h_{III}} \rangle =-0.008$, 
$ \langle {\rm h_{I} }\mid \rho \mid {\rm h_{IV} } \rangle =-0.071$. 
These four are the typical elements between nearest neighbor 
bond sites which can be easily understood from the view points 
of the  Wannier states \cite{geshi-etal.03}.

%The element 
%$ \langle {\rm h_{I} }\mid \rho \mid {\rm h_{II} } \rangle$ is that 
%forming a bonding orbit and 
%$ \langle {\rm h_{II}}\mid \rho \mid {\rm h_{III}} \rangle$ 
%is intra-atomic but inter-bond element. 
%These two are large components. 
%The other two  are relatively small components but 
%mainly contribute to inter-Wannier state interactions.  

When the dimension of the KS increases, 
the calculated values of off-diagonal elements of the 
density matrix  gradually approach to the exact values and saturate. 
Figure \ref{fig:off-density-matrix} shows the corresponding results for Si crystal with 512 atoms. 
In the present case they are saturated at around $\nu_{\rm K}=30$.
%Table. \ref{tab:off-density-matrix} 
The resultant convergent behavior and values are both excellent.
%The spatial behavior of the density matrix itself 
%will be discussed in more detail in Subsection \ref{subsec:Realsp}.

We note here that the convergence of the total energy could not be 
a unique measure of the convergence of the calculations. 
The convergence of the total energy is 
more rapid in comparison with that of the off-diagonal elements of the 
density matrix. 
In fact, the exact value of the band energy is $-5.082$~eV/electron 
and the calculated deviation from this 
is $+80,\ +23, \ +4,\ +1,\ +0$~meV/electron  
for $\nu_{\rm K}=7,\ 10,\ 20,\ 25, \ 30$, respectively.
%Since the error of less than 0.1eV per electron is acceptable 
%in many cases of the electronic structure calculations, 

%n   E{ave}  E_{ave}-E_{exa}(ev)
% 2 -4.63691 0.445644
% 3 -4.67248 0.410074
% 4 -4.68983 0.392724
% 5 -5.08204 0.000514
% 7 -5.00264 0.079914
%10 -5.05954 0.023014
%15 -5.07145 0.011104
%20 -5.07869 0.003864
%25 -5.08152 0.001034
%30 -5.08240 0.000154

It must be mentioned that, for the present covalent bonded systems, 
the generalized Wannier state can be reasonably reproduced by the 
first order perturbation theory of the sp$^3$ bonding orbitals and 
the Perturbative Order-N method is quite efficient
\cite{hoshi-fujiwara.00,hoshi-fujiwara.03}. 
The computational cost of the present KS method is less efficient 
in these systems.

%%%%%%%%%%%%%%%%%%
\subsection{Computational details and comparison with other methods}

In actual computations, we adopt the following procedure:\\
$[$i$]$ Generate the Krylov subspace defined by eq.~(\ref{eq:krylov1}) and 
generate eigen states within the KS by eq.~(\ref{eq:eigeneq}).\\
$[$ii$]$ Determine the chemical potential $\mu$ from the diagonal elements
of the density matrix  by using eq.~(\ref{chemPot}).   \\
$[$iii$]$ Calculate the off-diagonal elements of the density matrix 
through eqs.~(\ref{dm-KS}) and (\ref{eq:project3}).\\
$[$iv$]$ Calculate forces acting on each atom and move atoms.  \\
$[$v$]$ Return to the procedure $[$i$]$.

The computational time and memory size are mostly consumed 
in the part of generating the Krylov subspace. 
The computational cost of all other procedures is actually linearly 
proportional to the number of atoms. 
Furthermore, the only global quantity we use is 
the chemical potential $\mu$ 
and all other calculation is purely independent with respect to each starting vector.  
Therefore, the computational routine is parallelizable, 
and actually we made use 128 and 256 parallel processors 
with the Message Passing Interface (MPI) technique. 
%Because of the independent nature with respect to the basis set, 
%the efficiency of the parallel computation of the present method is perfect, XX.X\%. 
%We made use 128 and 256 parallel processors to simulate this calculations. 

Since the present method and the recursion method~\cite{Recursion_a,Recursion_b,haydock80} 
are both  based on the construction of the Krylov subspace, 
one might suppose that 
it were an extension of the recursion method. 
However, it is not the case.   
All calculations in the present method are based on the eigen values and eigen vectors 
in the Krylov subspace 
and one can calculate directly  off-diagonal elements of the density matrix. 
On the other hands, the recursion method is the way of 
calculating the diagonal Green's function 
in a form of the continued fraction. 
The discussion in the recursion method is always based 
on the diagonal elements of Green's functions  $G$.  
The proposed way to calculate the off-diagonal Green's function 
in the  recursion method may be~\cite{Recursion_a,Recursion_b,haydock80}  
\begin{eqnarray}
G_{ij}=\frac{1}{2}\Bigl( G_{i+j,i+j} - G_{i-j,i-j} \Bigr) ,
\end{eqnarray}
which needs a lot of computational  resources. 
The recursion method would recommend, in order to calculate 
the off-diagonal Green's function,  to use the 
%additional 
recurrence relation of the Green's function,~\cite{Ozaki99a,Ozaki00}
but it contains potential growth of a numerical rounding error.
%\begin{equation} G_{i,n+1}^{K(i)} = \left\{ G_{in}^{K(i)} (z-a_{n}) -
% G_{i,n-1}^{K(i)} b_{n-1} \right\} / b_{n},
%\end{equation}

The density matrix actually is given by the energy integration of the 
Green's function in the recursion method as 
\begin{equation} 
\rho_{ij} = -\frac{1}{\pi} \int_{-\infty}^{\infty} {\rm d}\varepsilon ~{\rm Im}G_{ij} (\varepsilon)
 f\Bigr(\frac{\varepsilon -\mu}{k_{\rm B}T}\Bigr) \ ,
\end{equation}
which causes a numerical error.  
The present method is completely free from above-mentioned 
difficulties in the recursion method. 
%
%On the contrary, in the present method, we use  eq.~(\ref{dm-KS}) 
%without any additional error. 
%
All calculations in the present method are based on the eigen values 
and eigen vectors 
in the Krylov subspace and one can calculate directly  
diagonal and off-diagonal elements of the density matrix simultaneously.

%%%%%%%%%%%%%%%%%%%%%%%%%%%%%%%%%%%%%%%%%%%%%%%%%

\section{Example : Results and Discussions for the Surface Reconstruction 
of Si $(001)$}
\label{sec:results}

In this section, we demonstrate how the electronic structure within the KS method 
gives the correct atomic structure. 
We show the results of molecular dynamics simulation 
of Si (001) surface reconstruction of a slab system of 1024 atoms. 
The essence of the quantum mechanical freedoms is the fact 
that sp$^3$-hybrid bonds are formed in the bulk region, 
but not on surfaces. 
Specifically surface atoms move to form asymmetric dimer
\cite{Chadi79,Ramstad-etal.95}. 
%The direct analysis of the Wannier states shows the drastic change 
%from the bulk sp$^3$ bonding to surface states. 
We will show the result of the present method and discuss the local electronic structure 
and the energy spectrum. 
%We will discuss the local electronic structure and the energy spectrum
%so as to analyse how the Krylov subspace method 
%reproduces the formation of the asymmetric dimer. 
%We will show that the asymmetry of the surface dimer is determined by 
%the local electronic structure. 
%through an illustration of its local electronic and cohesive properties. 
We also examine total energy difference for 
proposed three reconstructed configurations.

%%%%%%%%%%
\subsection{Tilt angle of surface dimers} 
\label{sec:tilt}

In ideal Si(001) surface, a pair of surface atoms has four electrons as dangling bonds. 
Two of them forms a $\sigma$-bonding state and a surface dimer appears. 
The other two electrons are directly related to the asymmetric geometry of the surface dimer. 
The Hilbert space for these electrons is restricted to the basis set orthogonal to the 
$\sigma$-bonding states and two back-bond states. 
If the four atomic orbitals, (s, p$_{x}$, p$_{y}$, p$_{z}$) per atom are considered, 
three freedoms are excluded by the orthogonality to the above three states. 
In the asymmetric dimer, the restricted basis set is given by an atomic basis of the $upper$ 
atom with a large s component and a relatively low energy level and the one of the $lower$ 
atom with a large p component and a relatively high energy level. 
Then the system can gain the energy, with the increase of s component, by charge transfer 
from the lower atom to the upper one. 
This mechanism is sometimes called `dehybridization' in the sense that the sp$^3$-hybridization 
is cancelled (See Ref.\cite{hoshi-fujiwara.03} and the references therein). 
In our previous work, we have observed a dynamical process of forming the asymmetric dimer, 
according to the above energy gain mechanism\cite{hoshi-fujiwara.03}. 
Therefore the present method should reproduce the above energy gain mechanism so as to 
reproduce the asymmetric dimer.

One of the factor to characterize the surface dimer 
is its tilt angle, $\theta$.
(See inset of Fig.\ref{fig:pdos}(a).)
%The ideal surface of covalent materials is energetically unstable because 
%the surface atoms have dangling bonds. 
%Then the surface atoms form a dimer but still a symmetric dimer, $\theta \sim 0$, is unstable. 
%One atom moves to the bulk (lower atom) 
%and the other moves to vacuum (upper atom) and 
%the pair forms an asymmetric dimer. 
%In an asymmetric dimer, the electron charge transfers from a lower atom 
%to an upper atom and the surface energy is lowered. 
Theoretical and experimental data of tilt angle $\theta$ are reviewed in ref.\protect\cite{Pollmann-etal.96}, 
and are ranging from 5\r{} to 19\r{}. 
The reported tilt angle by the exact diagonalization of the same 
tight-binding Hamiltonian is $\theta \sim$ 14\r{}\cite{Fu-etal.01}, 
while our result based on the KS method is $\theta=13.4$\r{} 
with the size of the KS $\nu_{\rm K}=30$. 
This result indicates that the present KS method 
extracts the essential character of the original Hamiltonian well. 
%Though the present tight-binding Hamiltonian is not tuned to the 
%surface calculation, the result is reasonably good. 
We will discuss in the next subsection that the asymmetric surface dimer is determined by 
the electronic states close to the chemical potential.

%Our result based on the KS method is $\theta=$13.4\r{\ }. 
%Though it is a little bit smaller than those of the reported values, the result is reasonable. 
%Here, we have chosen the size of the KS as $\nu_{\rm K}=30$ 
%with caveat stemming from sensitivity of surface dimer to the states energetically close 
%to the chemical potential. 
%%
%Just as the calculation with an improperly small size $\nu_{\rm K}$ leads to 
%an inaccurate value of the density matrix as mentioned in Subsection \ref{sec:rhoij-dep}, 
%the calculation with an improperly small $\nu_{\rm K}$ cannot reproduce 
%energy spectrum correctly, 
%in particular its profile close to the chemical potential. 

%Though a smaller size $\nu_{\rm K}$ is preferable for the computation 
%time of present method which is proportional to $\nu_{\rm K}$, 
%an improperly smaller size leads to unphysical results. 
%This is attributed to inaccurate value of the density matrix with 
%small size $\nu_{\rm K}$ calculation as mentioned previously. 
%Then at least $\nu_{\rm K} \sim 30 $ is required to simulate the system 
%and, hereafter we adopt $\nu_{\rm K} = 30$.

%we note here that the tilt angle is sensitive to the value of $\nu_{\rm K}$ 
%As mentioned in previous section, small size $\nu$ calculation leads to 
%inaccurate value of the density matrix. 

%%%%%%%%%%%%%%%%%

\subsection{Energy spectrum  and the  density of states}

While methods of density matrix may usually not provide an information about 
energy spectrum of electronic structure, 
the present method can do at the same time. 
To discuss the electronic spectra in the framework of the KS method, 
we introduce the Green's function $G_{ij}(\varepsilon)$ ;
\begin{eqnarray}
  G_{ij}(\varepsilon) = [(\varepsilon + {\rm i}\delta -H)^{-1}]_{ij} , 
\end{eqnarray}
where $\delta$ is an infinitesimally small positive number. 
Since the replacement for the density matrix (\ref{appr}) is guaranteed, 
the same replacement for the Green's function is also allowed;
\begin{eqnarray}
G_{ij}(\varepsilon) \Rightarrow G_{ij}^{K(i)}(\varepsilon) ,
\end{eqnarray} 
where the matrix elements of the Green's function in the KS is defined as
\begin{eqnarray}
G_{in}^{K(i)}(\varepsilon) 
=
 \sum_\alpha^{\nu_{\rm K}} \frac{C_{\alpha i}^* C_{\alpha n}}{\varepsilon + {\rm i}\delta -\varepsilon^{(i)}_{\alpha}} 
\label{eq:off-green-lanczos-vector1} \ . 
\end{eqnarray}
Actually the Green's function $G_{ij}(\varepsilon)$ can be calculated 
with the Green's function $G_{in}^{K(i)}(\varepsilon)$  in the KS as;
\begin{equation}
G_{ij}^{K(i)}(\varepsilon)  = \sum_{n}^{\nu_{\rm K}} G_{in}^{K(i)}(\varepsilon) 
                       \langle K_n^{(i)} \mid j \rangle .
\label{eq:off-green-lanczos-vector2}
\end{equation}
The equation (\ref{eq:off-green-lanczos-vector2}) is equivalent 
to (\ref{eq:project3}) and can be proven similarly by using the 
projection operator ${\hat P}^{K(i)}$.

%---------------------------------------
\begin{figure}[htbp]
\begin{center}
\resizebox{0.48\textwidth}{!}{
%  \includegraphics{./PDOS01.eps}
% original figure: ./SGI-JAERI/SLD24a1024pdos/figures/PDOS01.ps
%  \includegraphics{./DIMER01.eps}
% original figure: ./SGI-JAERI/SLD24a1024pdos/figures/DIMER01.ps
  \includegraphics{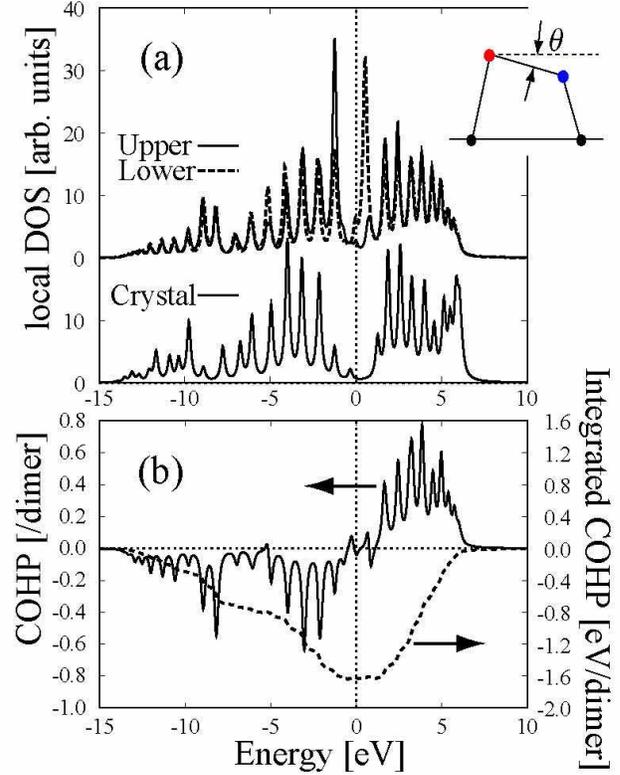}
% original figure: ./SGI-JAERI/SLD24a1024pdos/figures/DIMER01shift2.ps
}
\end{center}
\caption{ 
(a)Local density of states (lDOS) per atom for the system 
with asymmetric dimer and that for the system of crystal. 
Solid line (broken line) in upper panel represents an upper (lower) atom 
of the asymmetric dimer. 
(b) COHP and integrated COHP for the corresponding dimer. 
The energy zeroth both in (a) and (b) are common and is set to be the 
top of the occupied states in the bulk. 
In order to show the structure we introduce finite imaginary part, $\delta=0.136$eV, 
in the energy denominator of the Green function. 
The size of the reduced matrix is $\nu_{\rm K}=30$ 
and the temperature factor of the system in eq.~(\ref{dm-KS}) is $T=1580$ K (=0.136eV). 
The chemical potential is estimated as $\mu=$0.126eV. 
}
\label{fig:pdos}
\end{figure}
%---------------------------------------
%  \mu_{ori}  = 0.587836 eV
%  E_{vtop}   = 0.461392 eV
%  \mu_{shown}= 0.126444 eV
%
%
In order to single out the physical insight behind the asymmetric dimer, 
we calculate local density of states (lDOS) per atom 
of the system with reconstructed surface with dimer as shown in 
Fig.~\ref{fig:pdos}(a). 
The lDOS can be defined as 
%
%\begin{eqnarray} 
%n_{\bf R}(\varepsilon)
%&=&
%-\frac{1}{\pi} \sum_{\kappa}{\rm Im} G_{{\bf R}\kappa \;{\bf R}\kappa}(\varepsilon) 
%\\
%&=&
%-\sum_{\kappa} \mid \langle i \mid w^{(i)}_\alpha \rangle \mid^2 
%\delta(\varepsilon - \varepsilon^{(i)}_{\alpha}),
%\end{eqnarray} 
%%
%where ${\bf R}$ and $\kappa$ are the atomic site and orbitals. 
%
\begin{eqnarray} 
n_{I}(\varepsilon)
&=&
-\frac{1}{\pi} \sum_{\alpha}{\rm Im} G_{I \alpha , I \alpha}(\varepsilon) 
\\
&=&
\sum_{\alpha, \kappa}^{\nu_{\rm K}} \mid \langle I \alpha \mid w^{(I \alpha)}_\kappa \rangle \mid^2 
\delta(\varepsilon - \varepsilon^{(I \alpha)}_{\kappa}),
\end{eqnarray} 
where $I$ and $\alpha$ are the atomic site and orbitals, respectively,  
and $\kappa$ is suffix for eigen states of the KS. 
First of all, we see the lDOS of crystal. 
Because of the finite number of computed levels, $\nu_{\rm K}=30$, the shown lDOS has thirty spikes 
with weight factor $\mid \langle I \alpha \mid w^{(I \alpha)}_\kappa \rangle \mid^2 $
distributed from bottom to top of the band. 
%There are several sophisticated methods to smooth out these spiky structure in the 
%spectrum if it is necessary. 
Here we have introduced finite imaginary part, $\delta=0.136$eV ($10^{-2}$~Ryd), 
to smooth out these spiky structure. 
The calculated lDOS of crystal reproduces the gap that lies within $0 \sim 1$eV satisfactory. 
The lDOS of the deeper layer of the present slab system is similar to this and does not change before and after 
the surface reconstruction as it should be.  
In the lDOS for dimerized surface atoms, 
the lDOS of the upper (lower) atom has peak at $-1.25$ ($+0.54$)~eV 
in Fig.~\ref{fig:pdos}(a). 
% ori   : -0.789 (1.00683) eV 
% shift : -1.250 (0.54444) eV
The former (latter) peak corresponds to occupied (unoccupied) surface state 
and the difference of the spectra represents the electron charge transfer 
from the lower atom to the upper atom in the asymmetric dimer, 
as explained in \S \ref{sec:tilt}. 
In other words, the Krylov subspace method reproduces 
the electronic structure in the asymmetric dimer. 
%In other words, this energetical separation of the occupied and unoccupied surface state 
%is the origin of the asymmetric dimer. 
%
%Due to the atomic deformation in the asymmetric dimer, 
%an atomic (non-bonding) orbital of the upper atom gains the energy, 
%while another atomic orbital of the lower atom loses the energy.
%The energy gain mentioned above of the non-bonding orbital of the upper atom 
%is essential for the mechanism of forming the asymmetric dimer
%\cite{hoshi-fujiwara.03,Chadi79,Ramstad-etal.95}.
%In other words, 
%the energetical separation of the occupied and unoccupied surface state 
%should be reproduced  
%for the correct conclusion of the asymmetric dimer,
%as in the present calculation.

We note here about the two controlling parameters to reproduce the 
asymmetric dimer; the size of the KS, $\nu_{\rm K}$, 
and the temperature factor of the system, $T$. 
Both may affect the convergence speed in eq.~(\ref{eq:project3}) 
as well as the energy resolution of the simulation. 
The choice of $\nu_{\rm K}$ is important to reproduce the asymmetric dimer 
since the surface dimer reflects the electronic structure close to the chemical potential, 
in particular the occupied and unoccupied surface states. 
The size of the KS should be chosen so large that 
the profile of the surface states are well reproduced. 
Actually, the calculation with $\nu_{\rm K} < 20$ leads unstable value of $\theta$, 
for example, $\theta=0.2, 9.8, 14.5, 4.6$\r{\ } for  $\nu_{\rm K} = 15, 16, 17, 18$, respectively. 
While those with $\nu_{\rm K} > 25$ gives stable value, $13 \sim 14$\r{\ }. 
We have chosen $\nu_{\rm K} = 30$. 
%Hereafter we adopt $\nu_{\rm K} = 30$. 
%
%just as the calculation with an improperly small size $\nu_{\rm K}$ leads 
%to an inaccurate value of the density matrix as mentioned in Subsection 2.4, 
%
%It is contrastive with the density matrix representing whole property of 
%electronic structure from the bottom of the band to the chemical potential 
%as shown in eq.~(\ref{dm-KS}). 
%
%The convergence of $\theta$ to the choice of $\nu_{\rm K}$ reflects this and is 
%severer than those of the density matrix.  
%
The choice of $T$ is also important 
since the surface states are energetically close to the chemical potential. 
The temperature should be chosen so small that the occupied and unoccupied surface states 
are well separated energetically. 
%A temperature factor larger than this separation smears the structures 
%and results in symmetric dimer. 

%Also the formation of the asymmetric dimer creates the energy difference again 
%on the reconstructed surface. 
%[comment on temperature dependence]

%%%%%%%%%%%%%%%%%%%%%%%%%%%%%%%%%%%%%%%%%%%%%%%%%

In order to see the chemical bonding in condensed matters, we introduce the following quantity; 
\begin{equation}
C_{IJ} (\varepsilon) 
=
-\frac{1}{\pi} \sum_{\alpha,\beta}{\rm Im} G_{I \alpha, J \beta} (\varepsilon) H_{J \beta, I \alpha}. 
%C_{ij} (\varepsilon) 
%=
%-\frac{1}{\pi} \sum_{\alpha,\beta}{\rm Im} G_{i \alpha, j \beta} (\varepsilon) H_{j \beta, i \alpha}. 
\end{equation}
This is sometimes called the crystal orbital Hamiltonian populations (COHP)
\cite{dronskowski-blochl.93}.
The integration of this quantity gives cohesive energy 
from a pair of atoms just as the integration of local DOS gives occupation number. 
Actually, the total energy is decomposed into contributions of 
each atom pair as a sum of integration over the energy of $C_{IJ}$; 
\begin{eqnarray}
{\rm Tr}(\rho H) 
&=&
\sum_{I,J} \sum_{\alpha, \beta} \rho_{I \alpha, J \beta} H_{J \beta, I \alpha} 
\\
&=&
\sum_{I,J} \int_{-\infty}^{\varepsilon_{\rm F}} C_{IJ}(\varepsilon) {\rm d}\varepsilon. 
%{\rm Tr}(\rho H) 
%&=&
%\sum_{i,j} \sum_{\alpha, \beta} \rho_{i \alpha, j \beta} H_{j \beta, i \alpha} 
%\\
%&=&
%\sum_{i,j} \int_{-\infty}^{\varepsilon_{\rm F}} C_{ij}(\varepsilon) d\varepsilon. 
\end{eqnarray}
%
%Figure \ref{fig:pdos} shows the spectrum of calculated one in the KS with 
%a finite width of $\delta=0.136$eV and a spiky structure is seen. 
The analysis of the COHP and the integrated COHP shows where and how the bond formation 
stabilizes energetically the system. 
The COHP for the dangling bond pair (in ideal surface) is negligible (not zero), 
because interaction matrix element $H_{J\beta,I\alpha}$ within the dangling bond pair is very small 
due to a larger interatomic distance. 
Once an surface dimer is formed (though the atomic pair is the same), 
the COHP gives a finite value (Fig.~\ref{fig:pdos}(b)), 
because the interatomic distance is shortened and the interaction matrix element becomes finite. 
The integration of the COHP has its minimum almost at the chemical potential. 
This is a demonstration of the cohesive mechanism of covalent bonded materials. 
%The electron charge transfers from the lower atom to the upper atom, 
%then the dimer is stabilized energetically. 
%It is consistent with .....
%The energy spectrum of the COHP or the integrated COHP show where and how the bond formation 
%stabilizes energetically the system. 

%%%%%%%%%%
\subsection{Energy difference between different configurations of 
dimerized  ~{\rm Si} ${\rm (001)}$ surface}

The dimers may align on the Si (001) surface with three proposed 
reconstructed surface configurations, 
$(2 \times 1), (2 \times 2),$ and $(4 \times 2)$ 
\cite{Chadi79,Ramstad-etal.95}. 
Among them, the present calculation indicates that 
the $(4 \times 2)$ configuration is that of the lowest energy. 
%as suggested by experimental observation and the LDA calculations,~
%\cite{Ramstad-etal.95} .
The calculated energy differences from the $(4 \times 2)$ structure are 
86.7 meV/dimer for $E_{(2 \times 1)}$ and 0.3 meV/dimer for $E_{(2 \times 2)}$. 
%Though we used the tight-binding Hamiltonian, 
These values agree well with the exact calculation using the same Hamiltonian, 
$E_{(2 \times 1)}-E_{(4 \times 2)}=73.6$ meV/dimer and $E_{(2 \times 2)}-E_{(4 \times 2)}=1.2$ meV/dimer, 
respectively \cite{Fu-etal.01}. 
This shows that the numerical error with the KS method is small and 
the present method gives a satisfactory results in a fine energy scale 
with tight-binding Hamiltonian. 
On the other hand, we should comment that the tight-binding formulation itself can be the 
another origin of an error. 
In general, the energy scale in meV/atom is too fine to discuss in the present 
tight-binding Hamiltonian. 
An {\it ab initio} calculation gives 
$E_{(2 \times 1)}-E_{(4 \times 2)}=51 \pm$ 21 and 
$E_{(2 \times 2)}-E_{(4 \times 2)}=3 \pm$ 13 meV/dimer, respectively
\cite{Ramstad-etal.95}.
%We should use more sophisticated tight-binding Hamiltonian, 
%in order to discuss details about the dynamical change 
%between them at elevated temperatures.

%------------------------------------
%\begin{table}[htbp]
%\caption{
%Calculated energy difference among the different configurations of 
%the proposed Si(001) reconstructed surface in units of meV/dimer. 
%The energies are calculated by the KS method (present work), 
%an exact tight-binding (TB) calculation \cite{Fu-etal.01} with the same Hamiltonian, 
%and an {\it ab initio} calculation \cite{Ramstad-etal.95}.
%}
%\label{tab:t1}
%\begin{tabular}{cccc} \hline
%    & present calc. & exact TB & {\it ab initio} \\ \hline
%$E_{(2 \times 1)}-E_{(4 \times 2)}$ & 86.7 & 73.6 & 51 $\pm$ 21 \\
%$E_{(2 \times 2)}-E_{(4 \times 2)}$ &  0.3 &  1.2 &  3 $\pm$ 13 \\ \hline
%\end{tabular}
%\end{table}
%1024 Si atom ($8 \times 8 \times 16$ layer)
%$(2 \times 1)$
%$(2 \times 2)$ in phase domain
%$(4 \times 2)$ antiphase domain 
%
%------------------------------------

%%%%%%%%%%%%%%%%%%%%%%%%%%%%%%%%%%%%%%%%%%%%%%%%%
\section{Conclusions}
\label{sec:conclusion}

%1
In the present paper we presented a novel method using 
the Krylov subspace for the molecular dynamics simulation 
based on large-scale electronic structure calculation. 
%
%2
%The method provides the practical way by reducing 
By means of the reliable treatment of the reduced matrix deduced from 
the Krylov subspace method, 
the method provide an efficient and practical way to calculate the density matrix. 
%
%3
The method also provides a way to calculate the energy spectrum 
on the same standpoint as the density matrix. 
%
%4
As an example, the method is applied to the problem of the surface reconstruction 
of Si (001). 
%and we show the electronic process in the asymmetric dimer formation. 
%
We have pointed out through its analysis that the appropriate choice of 
the two controlling parameters, the size of the Krylov subspace and the temperature factor,
is important.
Both may affect the computational cost and the accuracy. 
Though the present calculation is just one example, 
it leads us to a general guiding principle 
in choosing the optimal values of the controlling parameters.

%computational cost

%We have pointed out that the temperature may affect 
%the convergence speed in eq (X),
%which means that 
%the computational cost may depend on 
%the choice of the temperature, 
%as well as the number of the bases in the Krylov subspace. 

%Since silicon has the band gap of about 1 eV and
%surface states appear between the energy gap,  
%the present discussion gives a typical energy resolution 
%for the reliable electronic structure calculations.

%5
%based on the density matrix formalism. 
%
%and give an extension of 
%methodology of the quantum mechanical molecular dynamics simulation. 
%
%The present KS method is alternative practical method, 
%without the generalized Wannier states. 
%

%6
In the present methodology 
the computational procedure of the density matrix, $\rho_{ij}$, is
independent for each atomic orbital, $i$, 
except the determination of the chemical potential, 
then the present method is very preferable for the parallel computation. 
%8
Moreover, this independency of the basis lead us 
a hybrid scheme within quantum mechanics \cite{hoshi-fujiwara.03}. 
In the hybrid scheme, 
the density matrix is decomposed into sub matrices 
and the sub matrices are determined by different methods. 
%Physical quantities represented as eq.~(\ref{eq:physica-quantity}) 
%are also decomposed accordingly and estimated separately. 
Molecular dynamics simulation with 10$^5$ atoms by 
the hybrid scheme between the present KS method 
and the perturbative Wannier state method is examined 
and will be published elsewhere.

%9
Since this newly developed method is a general theory for large matrices, 
the method is applicable for not only covalent bonded materials 
but also other systems like metal. 
The present KS method has a potentiality of wide applicability, even in non-Hermitian matrix, 
since the fundamental concept lies in the general linear algebra of large matrices.

%%%%%%%%%%%%%%%%%%%%%%%%%%%%%%%%%%%%%%%%%%%%%%%%%
\section*{Acknowledgment}

The authors  thank S.-L.Zhang and T. Sogabe (University of Tokyo) 
for the discussion about the Krylov subspace method. 
The valuable  advise about parallel computation 
by Yusaku Yamamoto (Nagoya University) is also grateful. 
Computation has been done 
at the Center for Promotion of Computational Science and Engineering (CCSE) of 
Japan Atomic Energy Research Institute (JAERI) 
and also partially carried out by use of the facilities of the 
Supercomputer Center, Institute for Solid State Physics, University of Tokyo. 
This work is financially supported by Grant-in-Aid 
from the Ministry of Education, Culture, Sports, Science and Technology 
and also 
by "Research and Development for Applying advanced 
Computational Science and Technology" 
of Japan Science and Technology Corporation.


\begin{thebibliography}{99}
\bibitem{NOTE-TB}
The tight-binding Hamiltonian
in the present carbon case
is given by tuning the energy difference between $s$ and $p$ atomic orbitals
$(\varepsilon_p - \varepsilon_s)$ 
to be the  half of that in the silicon case, 
according to the universal tendency of the tight-binding parameters.
~\cite{hoshi-fujiwara.00,hoshi-fujiwara.01}
\bibitem{NOTE-locality}
Our previous paper shows that
the Wannier state in carbon is more localized than that in silicon 
as the function of bond step \protect{\cite{hoshi-fujiwara.00,hoshi-fujiwara.01}}.
\end{thebibliography}
\end{document}